\documentclass[journal]{IEEEtran}
\usepackage{cite}
\usepackage{amsmath,amssymb,amsfonts,bm,color}
\usepackage[ruled, lined, linesnumbered, commentsnumbered, longend]{algorithm2e}
\usepackage{graphicx}
\SetKwProg{Init}{Initialization: }{}{}
\usepackage{textcomp}
\usepackage{wrapfig}
\usepackage{enumitem}
\usepackage{float}
\usepackage{graphicx}
\usepackage{subfigure}
\usepackage{multirow}
\usepackage{array}
\usepackage{tabularx}
\newcolumntype{L}[1]{>{\raggedright\arraybackslash}p{#1}}
\DeclareMathAlphabet\mathbfcal{OMS}{cmsy}{b}{n}
\begin{document}
	\title{Path Evolution Model for Endogenous Channel Digital Twin Towards 6G Wireless Networks}
    \author{Haoyu Wang, Zhi Sun, Shuangfeng Han, Xiaoyun Wang,  Shidong Zhou, and Zhaocheng Wang \vspace{-20pt}
		\thanks{Haoyu Wang, Zhi Sun, Shidong Zhou, and Zhaocheng Wang are with the Department of Electronic Engineering, Tsinghua University, Beijing 100084, China (e-mail: wanghy22@mails.tsinghua.edu.cn; zhisun@ieee.org; zhousd@tsinghua.edu.cn; zcwang@tsinghua.edu.cn).}
	\thanks{Shuangfeng Han and Xiaoyun Wang are with the China Mobile Research Institute, Beijing 100053, China. (e-mail: hanshuangfeng@chinamobile.com; wangxiaoyun@chinamobile.com)}
	\thanks{Corresponding Authors: Zhi Sun and Xiaoyun Wang}}
	\maketitle
			\begin{abstract}
			Massive Multiple-Input Multiple-Output (MIMO) is critical for boosting 6G wireless network capacity. Nevertheless, high-dimensional Channel State Information (CSI) acquisition becomes the bottleneck of massive MIMO system. Recently, Channel Digital Twin (CDT), which replicates physical entities in wireless channels, has been proposed, providing site-specific prior knowledge for CSI acquisition. However, external devices cannot always be integrated into existing communication systems, nor are they universally available across all scenarios. Moreover, the trained CDT model cannot be directly applied in new environments, which lacks environmental generalizability. To this end, Path Evolution Model (PEM) is proposed as an alternative CDT to reflect physical path evolutions from consecutive channel measurements. Compared to existing CDTs, PEM demonstrates virtues of full endogeneity, self-sustainability, and environmental generalizability. Firstly, PEM only requires existing channel measurements, which is free of other hardware devices and can be readily deployed. Secondly, self-sustaining maintenance of PEM can be achieved in dynamic channel by progressive updates. Thirdly, environmental generalizability can greatly reduce costs for large-scale deployment. To facilitate the implementation of PEM, an intelligent and light-weighted operation framework is firstly designed. Then, the environmental generalizability of PEM is rigorously analyzed, which can effectively address the distribution shift across environments. Extensive simulation results reveal that PEM can simultaneously achieve high-precision and low-overhead CSI acquisition, which can serve as a fundamental CDT for 6G wireless networks. 
            \end{abstract}	
	\vspace{-20pt}
\section{Introduction}
	\label{sec:intro}
	
Massive Multiple-Input Multiple-Output (MIMO) is pivotal for 6G wireless networks, including ultra-high spectral efficiency, massive access, and immersive sensing capabilities \cite{ieeenet_vision_Saad_2020}. To fully leverage large-scale antenna arrays, Channel State Information (CSI) knowledge is critical for beamforming and precoding operations in massive MIMO systems. However, CSI acquisition overhead also rapidly increases in large antenna arrays \cite{jstsp_overview_lu_2014}. Consequently, effective throughput will not consistently grow with the number of antennas. 

High-dimensional CSI acquisition becomes an inevitable challenge to meet promising 6G applications. To support numerous 6G applications with high throughput, there are two main requirements for CSI acquisition: (1) \textit{high CSI precision} to facilitate beamforming and precoding operations; (2) \textit{low acquisition overhead} for simultaneously serving numerous User Equipments (UEs).

\subsection{Channel Digital Twin}
To resolve the bottleneck of high-dimensional CSI acquisition, mainstream approaches directly utilize the measured CSI for prediction, which can be enhanced by deep learning \cite{chcom_cui_continuous_2024}. However, their channel acquisition accuracy dramatically relies on pilot densities, which cannot achieve high CSI precision with low pilot densities. Additionally, existing deep learning-related channel extrapolation and interpolation heavily depend on the distribution of the training dataset, which encounters the generalizability challenge when the distribution of wireless channel significantly shifts in the test environment.

Recently, Channel Digital Twin (CDT) has emerged as an effective tool for CSI acquisition. CDT endeavors to reflect physics entities in wireless channels with digital representations, which offers site-specific prior knowledge for CSI acquisition. Compared to the aforementioned pure pilot-based approaches, CDT can greatly reduce pilot overhead to attain high-precision CSI, which offers an additional dimension for CSI acquisition. Currently, there are two main types of CDTs. One is vision information, which provides geometric and object information to attain the properties of wireless channel \cite{commag_twin_Alkhateeb_2023, twc_computer_xu_2023}. Another is Channel Knowledge Map (CKM), which serves as a channel property database tagged with location \cite{comst_CKM_zeng_2024}. To infer wireless channel property (e.g., beam index and path parameters) with input data, deep learning is usually adopted in current CDTs. 

Despite the prior knowledge, current CDTs still encounter two challenges. On the one hand, current CDTs cannot be instantly integrated into existing communication systems due to their nature of exogeneity. Once the extrinsic input outside of communication systems (e.g., images from cameras and locations from GPS devices) is unavailable, the operation of current CDTs will be interrupted. For instance, the operation can be challenging for indoor UEs due to occlusion and weak indoor GPS signals. On the other hand, learning of current CDTs is environment-dependent. Explicitly, neural networks in current CDTs need to be retrained when operating in new environments, which induces a huge data collection and model retraining burden for large-scale deployment. 

In summary, CDT is an effective tool to enable superior applications for 6G wireless networks. However, when considering the aforementioned two challenges, the operation of current CDTs is still challenging. Thus, the key question is \textit{can we find a new type of CDT that is endogenous within communication systems and is environment-generalizable?}

\subsection{Proposed Path Evolution Model}
\label{subsec: solution}
\begin{figure}[t]
    \centering
    \includegraphics[width=0.48\textwidth]{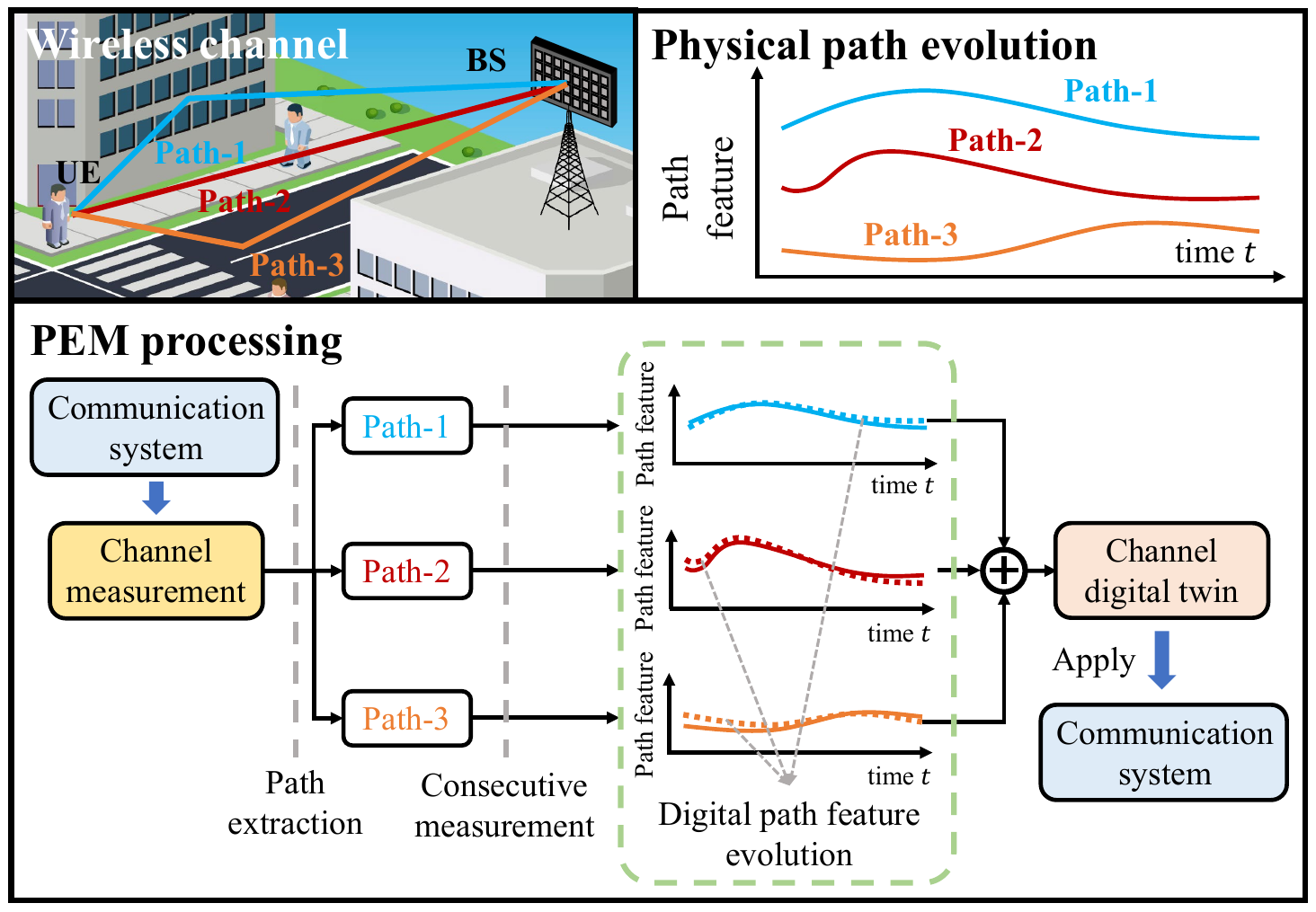}
    \vspace{-10pt}
    \caption{Structure of PEM, where digital replicas of physical path evolutions are created.}
    \label{fig: overview}
    % \vspace{-15pt}
\end{figure}

To this end, Path Evolution Model (PEM) is proposed as a novel CDT to meet the goals of \textit{intra-system endogeneity and environmental generalizability}. As shown in the top left of Fig.~\ref{fig: overview}, the wireless channel between the Base Station (BS) and UE is composited by multiple paths, which are basic physical entities under transmission, reflection, and scattering. Due to UE mobility, the path feature (e.g., path parameters and power distribution in the delay-angular domain) of each physical path consistently evolves with time, as shown in the top right of Fig.~\ref{fig: overview}. Since channel measurements are frequently conducted in existing communication systems, as shown at the bottom of Fig.~\ref{fig: overview}, the concept of PEM aims to extract the paths and then attain digital replicas of their feature temporal evolutions from the consecutive channel measurements. Then, CDT is yielded when digital replicas of all path evolutions are summed up. Thus, the constructed PEM can be applied to directly attain the desired path feature at future instants, which greatly reduces pilot overhead and enables the functionality of CDT. Firstly, the processing of PEM only requires existing channel measurements from communication systems, which is fully endogenous. Secondly, path feature evolutions in PEM are determined by universal electromagnetic (EM) propagation principles and regular user mobility. As a result, a generalizable path feature evolution can be assumed among dynamic environments, which facilitates environment-generalizability. 

\subsection{Advantages}
\label{subsec: adavantages}
The proposed PEM exhibits three inherent advantages over other existing CDTs: 
\begin{enumerate}
    \item {\bf Fully-endogenous}: PEM can be operated without introducing external hardware devices, which can be seamlessly integrated into existing communication systems.
    \item {\bf Self-sustaining}: PEM can effectively adapt to dynamic channels, where the digital path feature evolution can be equivalently viewed as a function of time and the explicit expression is controlled by historical path features. Since channel measurements are consecutive during the operation, the historical path features are progressively updated. 
    In this way, the constructed path feature evolution is consistently refreshed from the latest channel measurements, which can minimize the error between digital and physical path evolution in real time.
    \item {\bf Environment-generalizable}: With efficient learning, the neural network in PEM exhibits strong environmental generalizability. Such environmental generalizability greatly relieves the data collection and model retraining burden in new environments, which enables large-scale PEM deployment in 6G wireless networks. 
\end{enumerate}

\subsection{Our Contributions}
In this paper, we propose the concept of PEM, aiming to create digital replicas of physical path evolutions upon communication systems. Firstly, we propose an intelligent operation framework to efficiently construct, maintain, and apply PEM in dynamic wireless channels, which is compatible with existing communication systems and can be operated at a low cost. Secondly, the environment invariance of the target function of PEM is presented, which is fundamental to achieving environment generalizability. Thirdly, distribution shifts across environments are progressively addressed to enable environment generalizability for PEM. In the simulations, a good match can be found by comparing the physical and digital path evolutions, which validate the functionality of the proposed PEM. To justify the pilot overhead reduction capability of PEM, channel prediction under different time-domain pilot densities is investigated. Compared to CSI acquisition without CDT, both high-precision and low-overhead CSI acquisition can be achieved with PEM. Additionally, PEM exhibits endogeneity and strong environmental generalization capability. 

\section{Implementation of Path Evolution Model}
\subsection{PEM Framework} 

\begin{figure}[t!]
	\centering
	\includegraphics[width=0.48\textwidth]{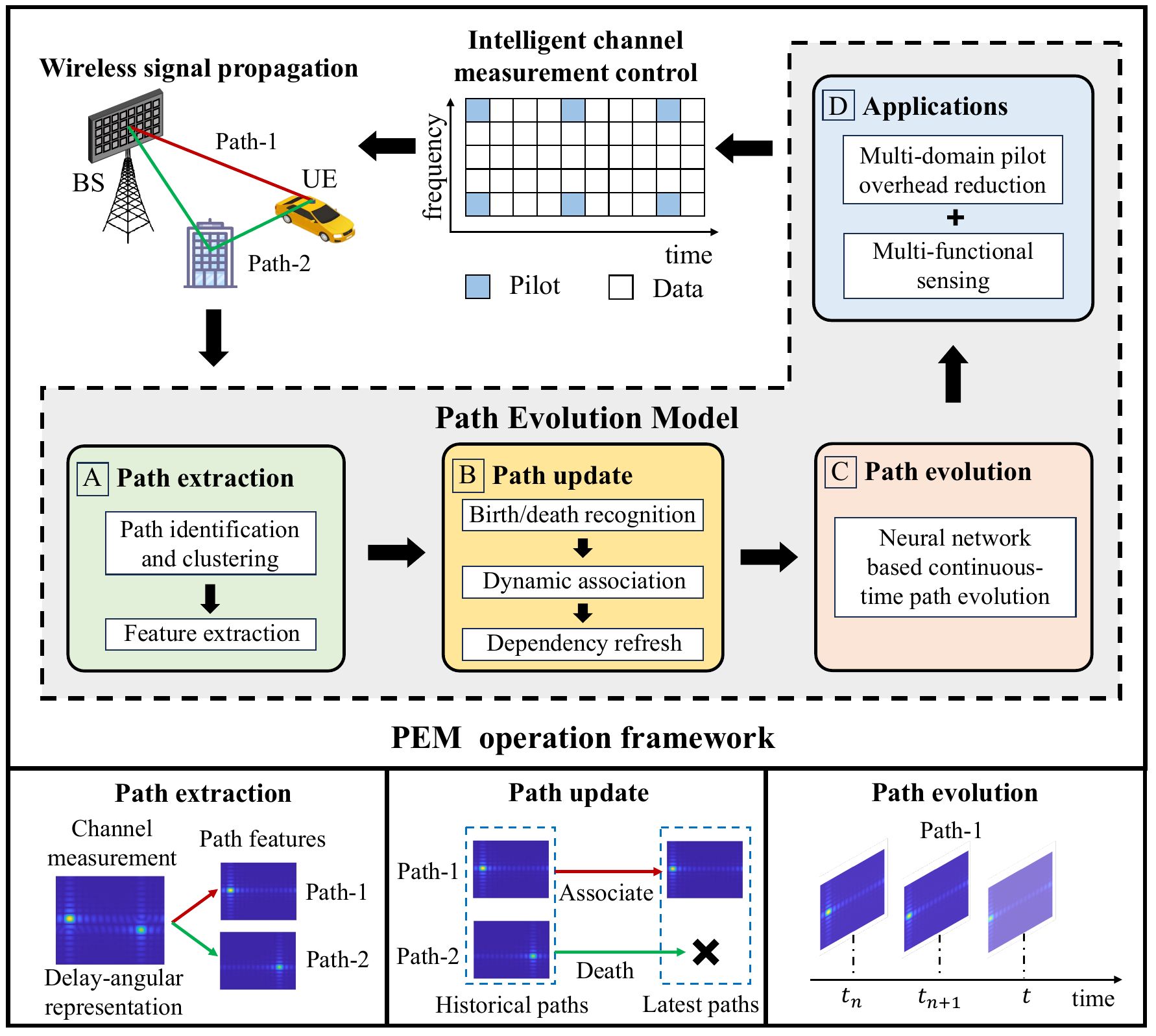} %
	\caption{PEM operation framework, where the path extraction, path update and path evolution steps are illustrated in the bottom part. }
	\label{fig: framework}
\end{figure}

The objective of the PEM operation framework is to construct, maintain, and apply digital replicas of path evolutions upon existing communication systems, which is illustrated in Fig.~\ref{fig: framework}. In existing communication systems, consecutive channel measurements can be attained by periodical pilot signals, e.g., Sounding Reference Signal (SRS). Then, the transmitted signals propagate from the UE to the BS through multiple paths. Thus, the channel measurements contain the information of each path. Due to the universal spatial consistencies \cite{tap_multiuser_Karstensen_2022}, path evolution exhibits obvious temporal correlation during UE mobility. Such consistencies enable continuous-time path evolution with discrete-time channel measurements. The procedure of PEM is demonstrated in the gray block of Fig.~\ref{fig: framework}, which contains three main steps:

\label{subsec: framework}

\begin{enumerate}
\item{\bf Path extraction}: The objective of path extraction is to attain the feature of each path from input channel measurements, as illustrated in the bottom left of Fig.~\ref{fig: framework}. For massive MIMO channel, each path can be characterized by the parameters of delay, Angle of Arrival (AoA) and power. Since the channel measurements are the sum of different paths, we first need to identify different paths with delay-angular representation or parameter estimation algorithms, e.g. SAGE and ESPRIT algorithms. In the sparse-scattering environment, path parameters can be adopted as path features. In the rich-scattering environment, multiple sub-paths with similar parameters form a path, which can be obtained via clustering. Then, path features can be represented as the power distribution in the delay-angular domain. The accuracy of path extraction is determined by the measurement resources, including bandwidth and the number of antennas.

\item{\bf Path update}: Path update aims to refresh the temporal path dependencies with the latest extracted path features, which is the key to enabling the self-sustainability of PEM. As shown in the bottom middle of Fig. 2, PEM can first recognize the birth/death status of paths by comparing the path feature distance with a pre-defined threshold. Thus, the proposed PEM is robust to the blockage scenarios by terminating the evolution of the disappearing Line of Sight (LoS) path. Next, the remaining latest path features can be associated with the surviving historical paths based on path feature distances, which can be solved by bipartite matching. Considering non-linearity in path feature evolutions, neural network is adopted to represent path feature temporal dependencies. The associated path features at the current instant are fed into the neural network to update the hidden states of each path, which refreshes path feature temporal dependencies. 

\item{\bf Path evolution}: The goal of path evolution is to attain the feature of each path at any desired instant via the updated temporal path feature dependencies, which is vital for real-time applications. The rationale lies in the fact that path features continuously vary between two adjacent update instants. Hereby, the non-linear path evolution is achieved by a well-trained neural network. With the previous path extraction step, the neural network for path evolution exhibits robust generalization capabilities across environments. Analysis of environmental generalizability is detailed in Sec.~\ref{subsec: addressing distribution shift advantage} and \ref{subsec: addressing distribution shift}. 
\end{enumerate}

Our proposed PEM operation framework is light-weighted and can be realized at a low cost. Firstly, PEM can avoid additional hardware devices and achieve protocol compatibility with existing communication systems. Secondly, considering the average lifetime of paths in usual dynamic scenarios, the update period in PEM processing can be set in the scale of 0.1s, which is far longer than the typical slot length and makes the pilot overhead affordable. Thirdly, the historical path feature sequence can be encoded in the hidden states of the neural network of PEM. Thus, PEM only needs to retain the extracted path feature at the current instant, which makes the memory occupation affordable.

\begin{figure*}[t]
		\centering		
        \includegraphics[width=0.95\textwidth]{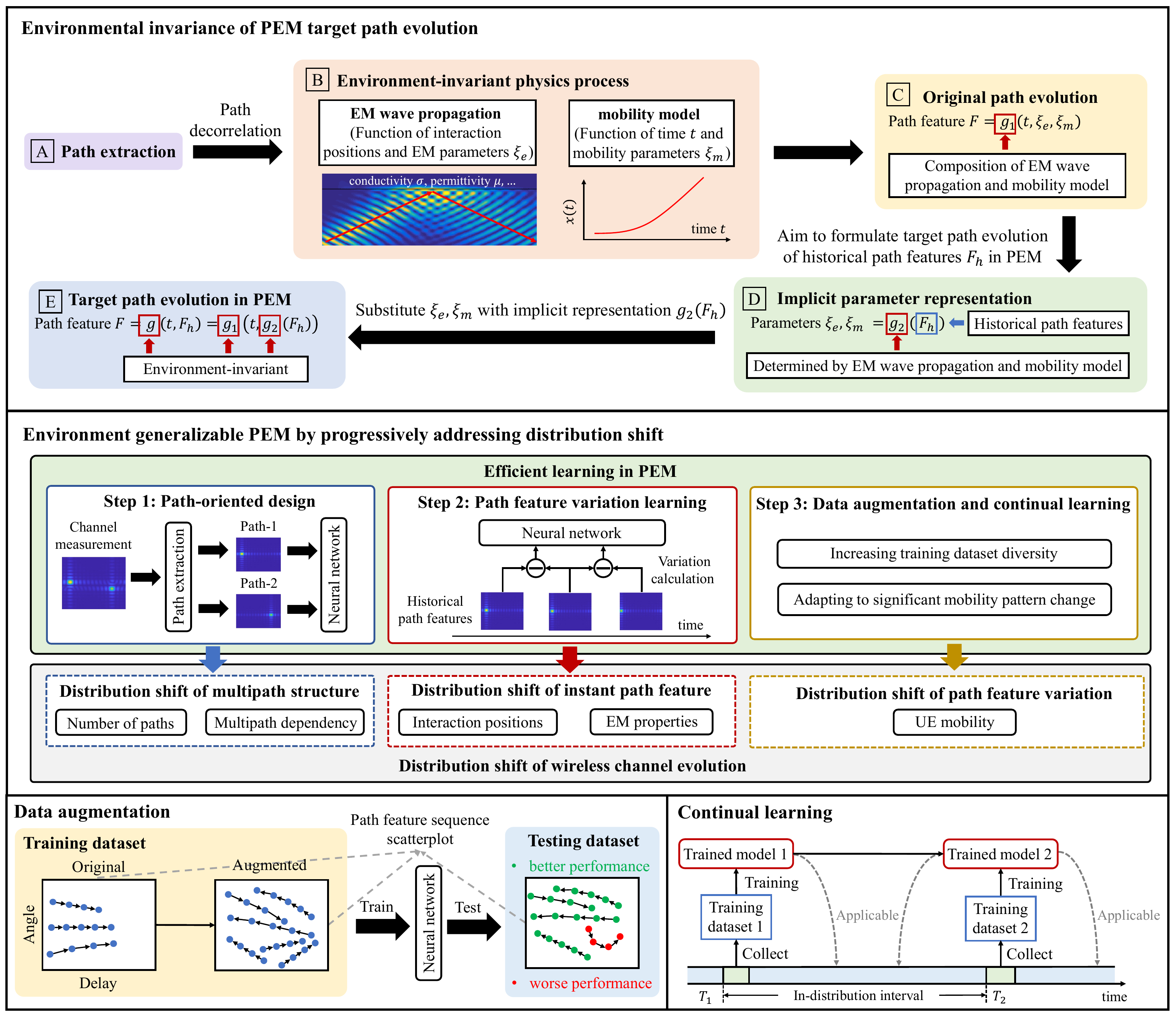}
		% \vspace{-5pt}
		\caption{Analysis of environment-invariant target function in PEM is shown on the top. Then, an overview environment generalizable PEM is illustrated in the middle, which progressively addresses the distribution shift. Next, data augmentation and continual learning that address the varying user mobility are detailed at the bottom.}
        \label{fig: generalization}
		% \label{fig: generalization analysis}
		% \vspace{-15pt}
\end{figure*}
\subsection{Environment Invariance of Target Function}
\label{subsec: addressing distribution shift advantage}

The invariance of target path evolutions to historical path features and time is fundamental for the PEM generalization capability. From block A to block E on the top of Fig.~\ref{fig: generalization}, analysis of invariant target path evolution of PEM is unfolded in a logically progressive manner below.
\begin{enumerate}
    \item \textbf{Block A}: To begin with, we can focus on the generalizability in the feature evolution of a single path based on the path extraction step in the operation framework. The evolution of multiple paths in wireless channels includes two aspects: the evolution of each single path and the correlation among different paths. Through path extraction, the environment-related path correlation can be removed. 
    \item \textbf{Block B}: Subsequently, the evolution of a single path originates from two physics processes, EM wave propagation and mobility. EM wave propagation (e.g., transmission and reflection) can be formulated as a function of interaction positions and EM parameters (e.g., conductivity and permittivity) along a path, which is universal among environments. The user mobility model is calculated based on kinematics, which can be parameterized by initial conditions, speed limit, and topology nodes. Without loss of generality, mobility model among environments can be unified \cite{comst_mobility_Harri_2009}. 
    \item \textbf{Block C}: Thereafter, original path evolution can be formulated as a function of time, EM, and mobility parameters based on the EM wave propagation laws and mobility model. 
    \item \textbf{Block D}: Then, input in the original path evolution should be reformulated as historical path features to facilitate PEM. With sufficient measurements and Signal-to-Noise Ratio (SNR), an auxiliary implicit expression of EM and mobility parameters can be attained from historical path features \cite{twc_electromagnetic_jiang_2024}. Since the implicit expression is determined by the EM propagation laws and mobility model, it is environment-invariant as well. 
    \item \textbf{Block E}: Finally, due to the invariance of EM wave propagation laws and mobility model, the environment-invariant target path evolution is yielded when the EM and mobility parameters are substituted by their implicit expressions.
\end{enumerate}

\subsection{Addressing Distribution Shift Across Environments}
\label{subsec: addressing distribution shift}
Based on the invariance of the target function, distribution shifts across environments should be addressed to enable environment generalizability for PEM. The distribution shift of wireless channel evolution during user mobility is composed of three factors: (1) the distribution shift of multipath structure; (2) the distribution shift of instant path feature; (3) the distribution shift of path feature variation. Firstly, the multipath structure includes the number of paths and the dependency among paths. Due to the diversity of shape, deployment, and EM properties of objects in wireless channels, the distribution of multipath structure significantly shifts across different environments. Secondly, the distribution of the instant path feature is determined by the interaction points along a path. Due to the variations of object configurations and user positions, the distribution of instant path feature diverges in different environments as well. Thirdly, path feature variation is caused by user mobility, where distribution shift occurs when user mobility pattern changes.

The proposed PEM can progressively address the aforementioned three types of distribution shift, which is shown in the middle of Fig.~\ref{fig: generalization}. Firstly, the neural network in PEM is designed in a path-oriented manner. Explicitly, the neural network individually realizes the evolution of each extracted path, which tackles the distribution shift of the multipath structure. Secondly, path feature variation learning is adopted. Hereby, historical path feature variations are calculated and input into the neural network, which predicts path feature variation in future instants. Then, the predicted path feature can be obtained by adding the predicted path feature variation and the latest historical path feature. For the LoS path and specular reflection paths, the variation of path features weakly correlates with the instant path feature, which largely mitigates the distribution shift of the instant path feature. Thirdly, data augmentation and continual learning are adopted to address the distribution shift of user mobility pattern. Compared to the object deployment and materials changes, user mobility exhibits a much lower variability across environments, which facilitates environment generalization. As shown in the bottom left of Fig.~\ref{fig: generalization}, data augmentation directly manipulates the training samples, which can enhance the diversity of the training dataset and is free of training samples from the unseen environment \cite{tkde_generalizing_wang_2023}. Typical data augmentation includes rotation and scaling, which benefits the generalization to new environments with different speed ranges and orientations. As depicted in the bottom right of Fig.~\ref{fig: generalization}, continual learning can be leveraged to adapt to the variation of user mobility pattern, which will be triggered when a large amount of out-of-distribution mobile users occur. Note that the in-distribution interval of user mobility pattern is far greater than the operation duration of a single user, data collection and model update in continual learning are tractable. Meanwhile, since the memories of seen environments can be retained under a continual learning framework \cite{tpami_continual_wang_2024}, the generalizability to new environments in PEM can also be gradually increased.

\section{Applications of Path Evolution Model}
\begin{figure}[t]
		\centering
		\includegraphics[width=0.5\textwidth]{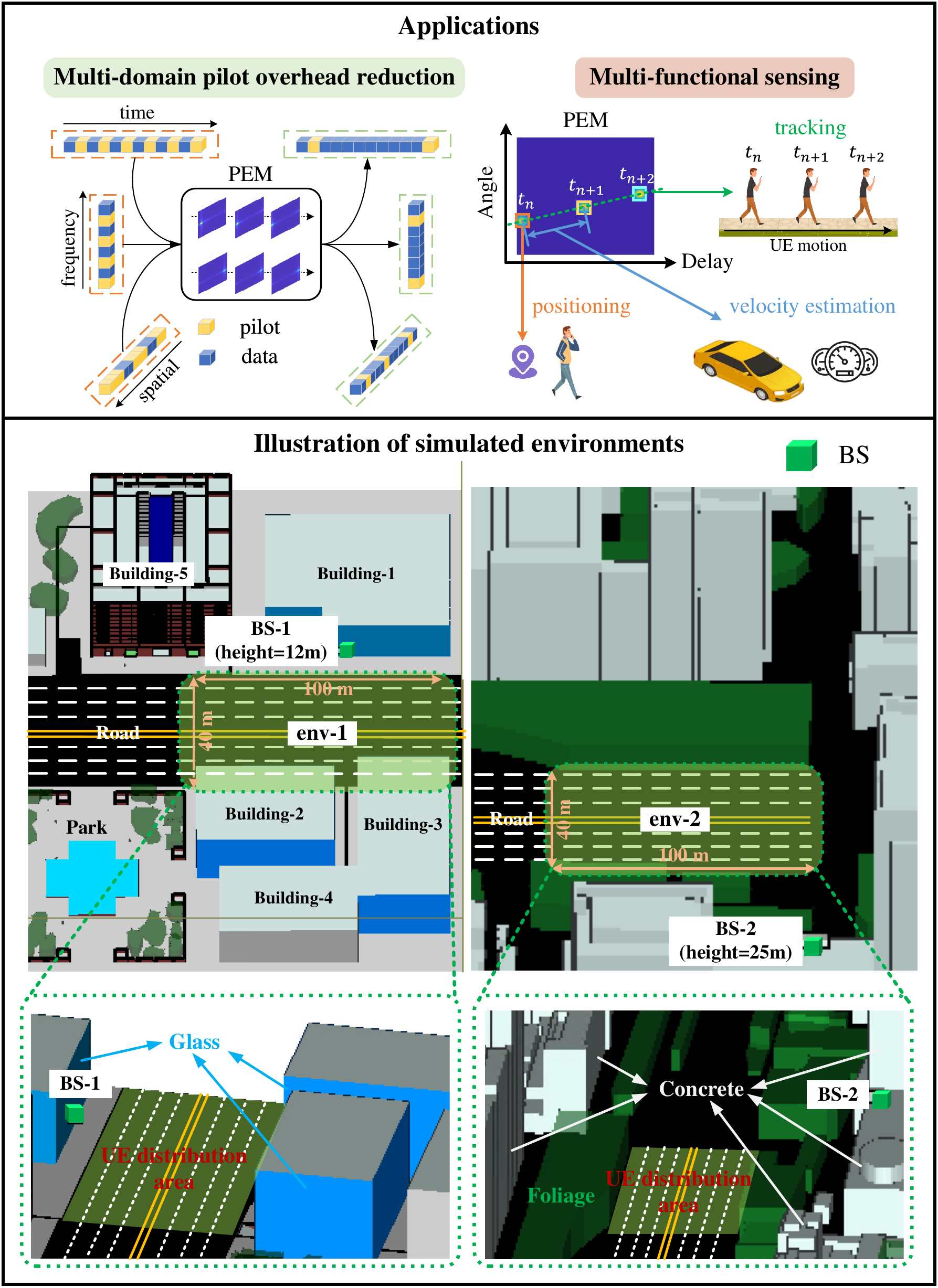}
		\vspace{-15pt}
		\caption{Illustration of PEM applications (top), along with simulation scenario in Wireless Insite (bottom).}
		\label{fig: sim env}
		\vspace{-10pt}
\end{figure}

Once PEM is constructed and maintained, it can be applied in real time. As shown in the top of Fig.~\ref{fig: sim env}, two main applications of PEM can be categorized:  
\begin{itemize}
    \item {\bf Multi-domain pilot overhead reduction:} With temporal dependencies of path feature in PEM, frequency/time-domain ambiguity \cite{Richards2022} for channel acquisition under low pilot densities can be tackled. Thus, pilot overhead in the frequency/time-domain can be greatly reduced below channel coherence bandwidth and coherence time limitations. Additionally, beam prediction via path evolution can reduce pilot overhead in spatial-domain \cite{wc_deep_ma_2023}. 
    \item {\bf Multi-functional sensing:} Note that explicit delay and angular information in path features characterize the geometrical relationship among BS, UE, and scatters. Instant path features can be used for positioning and the shift of path features can be leveraged for velocity estimation. Meanwhile, with path trajectories maintained in PEM, user tracking can be achieved as well. 
\end{itemize} 

The high-precision and low-overhead CSI acquisition capability of PEM is investigated by the following simulations. 

\subsection{Simulation Setup}
\label{subsec: setup}
In the simulations, the precise ray-tracer Wireless Insite is adopted to generate CSI data. As shown in the bottom of Fig.~4, street scenarios are considered, where mobile users are distributed in env-1 and env-2. Explicitly, env-1 and env-2 exhibit distinct geometric layouts, including building layouts, BS heights, and user positions relative to BS. Additionally, different types of materials are adopted in the buildings in env-1 and env-2. Mobility of users is generated by SUMO simulator \cite{ITSC_microscpoc_Lopez_2018}, where speed limits are set as 20 m/s and 30 m/s in env-1 and env-2, respectively. The communication systems operate at a 6 GHz carrier frequency, where the bandwidth is set as 100 MHz and the number of subcarriers is 138. Uniform Planar Arrays (UPA) with shape $16\times16$ are both equipped in BS-1 and BS-2. The SNR of channel measurement is set as 10 dB. For the neural network in PEM and deep learning-based baseline, a training dataset with 1600 samples is collected from env-1, which is free of training samples in env-2. Data augmentation in Sec.~\ref{subsec: addressing distribution shift} is adopted during the training of PEM. 

\begin{figure*}[t]
		\centering		
        \includegraphics[width=0.9\textwidth]{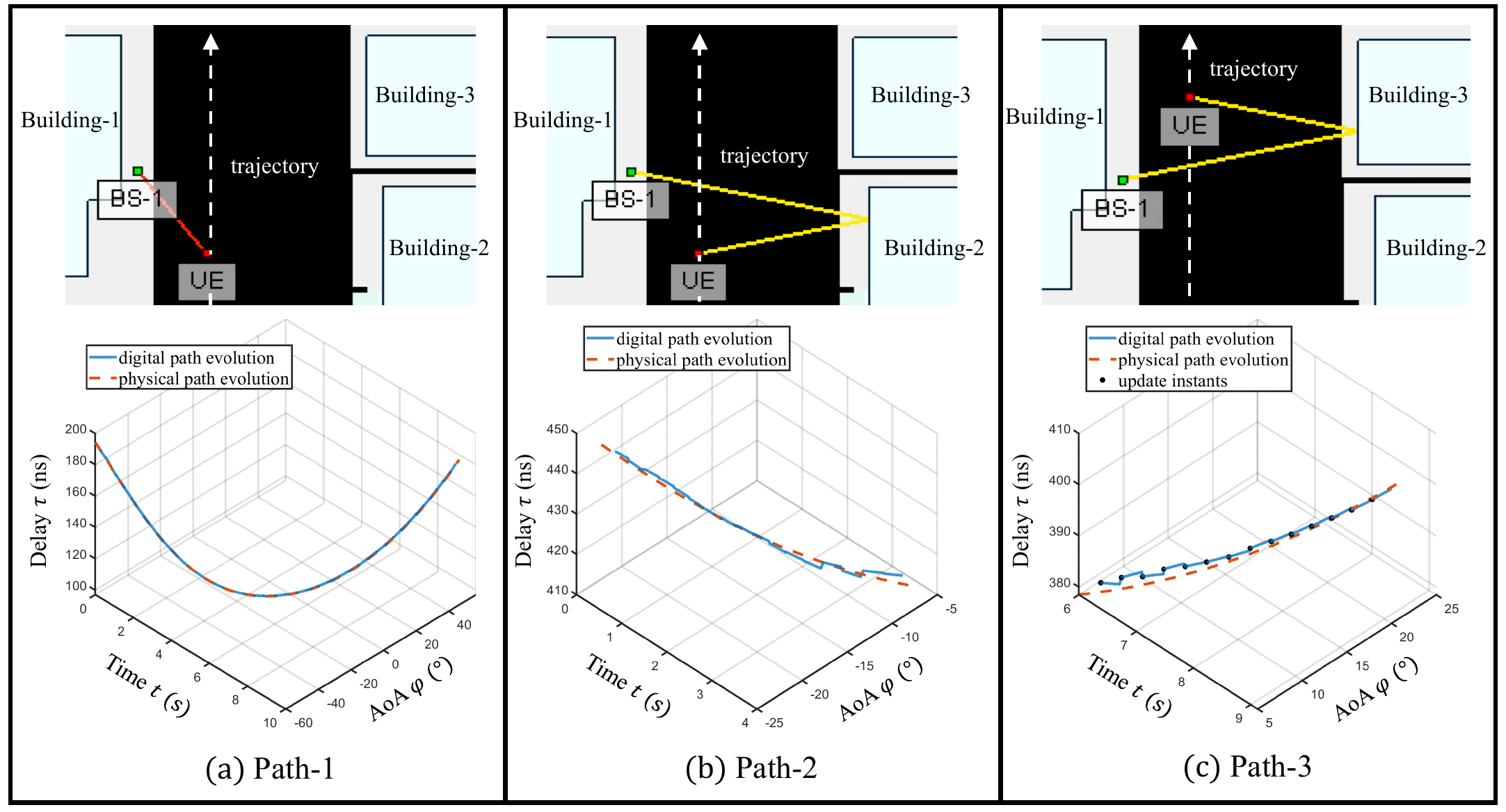}
		% \vspace{-5pt}
		\caption{An example of path evolution during UE mobility in env-1. Three physical paths are illustrated with a top view on the top side. Delay and AoA evolution of these paths are plotted on the bottom side: (a) path-1: LoS path; (b) path-2: reflection path from building-2; (c) path-3: reflection path from building-3.}
        \label{fig: visualization}
		% \label{fig: generalization analysis}
		% \vspace{-15pt}
\end{figure*}

\subsection{PEM Functionality}
Firstly, we evaluate the functionality of PEM by comparing the physical and digital path evolutions, which is illustrated in Fig.~\ref{fig: visualization}. Here, evolutions of three typical paths in env-1 are taken as examples, including one LoS path and two reflection paths. During UE mobility, it can be found that the digital evolutions of the three paths can accurately match their physical counterparts. Moreover, PEM can recognize the death of path-2 and the birth of path-3 when passing from building-2 to building-3. Thus, PEM can provide precise prior knowledge for CSI acquisition in real time. Additionally, the evolution error of path 3 in Fig.~\ref{fig: visualization}(c) can be progressively reduced with the updated channel measurements, which exhibits the self-sustainability of PEM.
\subsection{Time-domain Pilot Overhead Reduction}
\label{subsec: case 1}
In this subsection, we aim to leverage PEM to reduce pilot overhead in time-domain by increasing the SRS period. With high channel dynamics and a large SRS period, continuous time-domain channel prediction is vital to guarantee seamless high spectral efficiency. Hereby, tensor neural Ordinary Differential Equation (ODE) channel prediction network \cite{chcom_cui_continuous_2024} and model-based prediction \cite{twc_partial_twc_2022} is adopted as a continuous CSI prediction baseline without CDT, where the length of the historical CSI sequence is 8. Normalized Mean Square Error (NMSE) of channel prediction schemes under different SRS periods is shown in Fig.~\ref{fig: time domain}. It can be found that the NMSE of PEM is low and nearly remains unchanged when time-domain pilot overhead is reduced by 5 times, which benefits from the real-time prior knowledge of path feature evolution. Caused by the time-domain ambiguity, the NMSE of the baselines obviously increases when the SRS period is large. Thus, it is evident that PEM can simultaneously enable low-overhead and high-precision CSI acquisition. 

\subsection{Discussion over Generalization and Endogeneity}
\label{subsec: discussion}
\begin{figure}[!t]
\centering
\vspace{-15pt}
\includegraphics[width=0.45\textwidth]{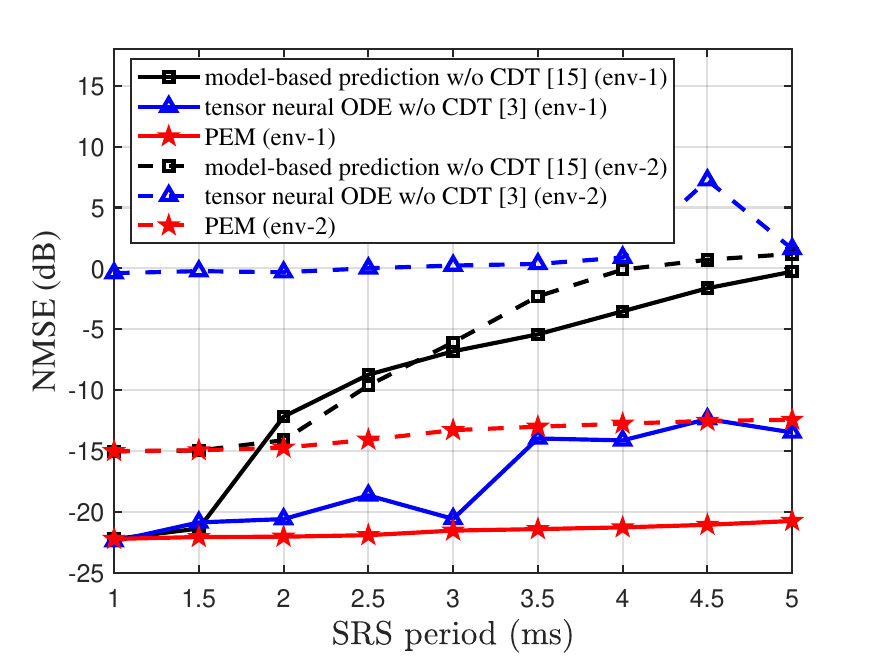}
\vspace{-5pt}
\caption{Time-domain channel prediction under different SRS periods.}
\vspace{-10pt}
\label{fig: time domain}
\end{figure}

Intra-system endogeneity and environment generalization can be achieved in PEM. As depicted in Fig.~\ref{fig: time domain}, the channel prediction performance of PEM is held in the unseen env-2, which is achieved without any external device. Note that 20\% of the user distribution area is blocked in env-2. Thus, the simulation results reveal that the PEM is robust to blockage. Conversely, the channel acquisition performance of the deep learning-based baseline is obviously degraded when tested in the unseen env-2, which shows the lack of generalizability. 

\section{Conclusions and Future Research}
CDT is a new paradigm for efficient CSI acquisition in 6G wireless networks. In this paper, PEM is introduced as a new type of CDT, which is designed to reflect path evolutions in the physical world. Compared to existing vision information-based and CKM-based CDTs, PEM possesses three key advantages of full endogeneity, self-sustainability, and environmental generalizability, which are vital to achieving low operation and deployment costs. To facilitate PEM, an intelligent operation framework is proposed, which is light-weighted and compatible with existing massive MIMO systems. The environment generalizability of PEM is rigorously analyzed, which is realized by progressively addressing the distribution shift. Based on extensive simulations, we validate that high-precision and low-overhead CSI acquisition among dynamic environments can be effectively achieved by PEM, which can enable large system capacity for 6G wireless networks. 

Although great potential and feasibility have been proved to adopt PEM for efficient CSI acquisition, its effectiveness relies on the specific dynamic nature of the wireless channel and the performance to process the input channel measurements during the operation. Hence, the future research directions span from environment-aware operation design to efficient data processing. Firstly, to optimize the operation performance of PEM, the environment-aware update period should be designed by intelligent decision-making. Secondly, the accuracy of path extraction in PEM is affected by the measurement resources. With limited measurement resources, how to achieve accurate path extraction in a rich multipath environment needs to be resolved. Thirdly, PEM can tackle channel measurement noise with extrinsic information. However, the discrepancies among data modalities bring challenges to multi-modal data fusion. Thus, a multi-modal data fusion framework that fully employs specific physics meanings of extrinsic information requires further investigation. Fourthly, in complex environments with diffusion paths and moving scatters, the distribution shift of the path features evolution becomes more complicated. Therefore, efficient processing algorithms in PEM based on a more comprehensive distribution shift model merit a deeper exploration.

\section{Acknowledgement}
This work was supported by the National Key R\&D Program of China under Grant 2022YFB2902004.

\bibliographystyle{IEEEtran}
\bibliography{commag}
    \begin{IEEEbiographynophoto}{Haoyu Wang} [S] (wanghy22@mails.tsinghua.edu.cn)
        received his B.S. degree from Tsinghua University in 2022, where he is currently working toward the Ph.D. degree with the Department of Electronic Engineering, Tsinghua University. His research interests lie in machine learning for communications and channel digital twin.  
    \end{IEEEbiographynophoto}
    \vspace{-36pt}
	\begin{IEEEbiographynophoto}{Zhi Sun} [SM] (zhisun@ieee.org)
		 received his Ph.D. degree from Georgia Institute of Technology in 2011. Currently he is a tenured Associate Professor at Tsinghua University, Beijing, China, which he joined in 2021. Prior to that, he was a tenured Associate Professor at University at Buffalo, the State University of New York, USA, which he joined in 2012 as an Assistant Professor. He received the US NSF CAREER Award in 2017. Zhi Sun has served as the editor for IEEE Transactions on Mobile Computing, IEEE Transactions on Wireless Communications, and Computer Networks (Elsevier). His research interests lie in wireless communications and networking. 
	\end{IEEEbiographynophoto}
    \vspace{-36pt}
    \begin{IEEEbiographynophoto}{Shuangfeng Han} [SM] (hanshuangfeng@chinamobile.com)
    received his Ph.D. degree in Electronic Engineering from Tsinghua University, in 2006. He is a Principal Researcher of China Mobile Research Institute. His research interests include massive MIMO and wireless artificial intelligence. He is an IEEE Senior Member, and is serving on the editorial board of IEEE Commun. Mag.
    \end{IEEEbiographynophoto}
    \vspace{-36pt}
    \begin{IEEEbiographynophoto}{Xiaoyun Wang}(wangxiaoyun@chinamobile.com) is the Chief Scientist and vice CTO of China Mobile. Her research interests include technology strategy, system architecture, and networking technology. She is the recipient of multiple National Science and Technology Progress Awards.
    \end{IEEEbiographynophoto}
    \vspace{-36pt}
    \begin{IEEEbiographynophoto}{Shidong Zhou} [M] (zhousd@tsinghua.edu.cn) received the B.S. and M.S. degrees from Southeast University, Nanjing, China, in 1991 and 1994, respectively, and the Ph.D. degree from Tsinghua University, Beijing, China, in 1998. He is currently a professor with the Department of Electronic Engineering, Tsinghua University. He was involved in the Program for New Century Excellent Talents in University 2005 (Ministry of Education). His research interest lies in wireless transmission techniques, including distributed wireless communication systems, channel sounding and modeling, coordination of communication, control and computing, and application in future mobile communications. He received the Special Award of the 2016 National Prize for Progress in Science and Technology.
    \end{IEEEbiographynophoto}
    \vspace{-36pt}
    \begin{IEEEbiographynophoto}{Zhaocheng Wang} [F] (wangzc@tsinghua.edu.cn) received his B.S., M.S., and Ph.D. degrees from Tsinghua University, in 1991, 1993, and 1996, respectively. From 1996 to 1997, he was a Post Doctoral Fellow with Nanyang Technological University, Singapore. From 1997 to 2009, he was a Research Engineer/Senior Engineer with OKI Techno Centre Pte. Ltd., Singapore. From 1999 to 2009, he was a Senior Engineer/Principal Engineer with Sony Deutschland GmbH, Germany. Since 2009, he has been a Professor with Department of Electronic Engineering, Tsinghua University. He was a recipient of IEEE Scott Helt Memorial Award, IET Premium Award, IEEE ComSoc Asia-Pacific Outstanding Paper Award, and IEEE ComSoc Leonard G. Abraham Prize.
    \end{IEEEbiographynophoto}
\end{document}